\documentclass[prb,preprint,aps]{revtex4}
\usepackage{amsmath}
\usepackage{graphicx}
\begin{document}
\title{Quantum metamaterials: Electromagnetic waves in a Josephson qubit line}
%
%
\author{A.L. Rakhmanov$^{1,2}$}
\author{A.M. Zagoskin$^{1,3,4}$}
\author{Sergey Savel'ev$^{1,3}$}
\author{Franco Nori$^{1,5}$}

\affiliation{(1) Frontier Research System, The Institute of Physical
and Chemical Research (RIKEN), Wako-shi, Saitama, 351-0198, Japan}
\affiliation{(2) Institute for Theoretical and Applied Electrodynamics
RAS, 125412 Moscow, Russia}
\affiliation{(3) Department of Physics, Loughborough University,
Loughborough LE11 3TU, United Kingdom}
\affiliation{(4) Physics and Astronomy Dept., The University of British Columbia, Vancouver, B.C., V6T 1Z1, Canada}
\affiliation{(5) Department of Physics, Center for Theoretical Physics, Applied Physics Program, Center for the Study of Complex Systems, University of
Michigan, Ann Arbor, MI 48109-1040, USA}

\date{\today}

\begin{abstract}
 We consider the propagation of a classical electromagnetic wave through a transmission line, formed by identical superconducting charge qubits inside a superconducting resonator. Since the  qubits can be in a coherent superposition of quantum states, we show that such a system demonstrates interesting new effects, such as a ``breathing'' photonic crystal with an oscillating bandgap,  and a ``quantum Archimedean screw'' that transports, at an arbitrary controlled velocity, Josephson plasma waves through the transmission line. The key ingredient of these effects is that the optical properties of the Josephson transmission line are controlled by the quantum coherent state of the qubits.
\end{abstract}
 \maketitle

\section{Introduction}
The development of superconducting electronics  now allows the observation of quantum behavior, such as the coherent superposition of different macroscopic states in meso- and macroscopic devices~\cite{You2005a,Wendin2006}.
While efforts in this field are now mainly directed at the development of superconducting quantum bits as elements of eventual quantum computers, there are other interesting possibilities opened by the existence of such large, controllable, quantum coherent circuit elements. In particular,  analogies with cavity QED have led to interesting  theoretical and experimental results~\cite{Blais2003,you:2003a,you:2003bb,Wallraff2004,Zagoskin2004,You2007}. There, qubits play the role of artificial atoms, and  high-quality superconducting resonance circuits mimic optical cavities. Differences include that instead of having a stream of identical atoms moving through the cavity, the state of a single qubit, permanently coupled to the resonator, could be periodically changed. In Ref.~\onlinecite{Hauss2007}  results of an earlier experiment~\cite{Ilichev2003} were  considered from the point of view of lasing in such a system.

Another recent surge of interest in the electrodynamics of Josephson
Junction arrays is related to THz electromagnetic waves propagating in such systems. THz waves are
important for applications, but are hardly controllable for both
optical and electronic devices. Thus, Josephson structures might
be of potential importance for miniature THz generators, filters,
detectors, and wave guides~\cite{thz}.

In this paper, instead of considering a single or a few qubits, we
investigate the behaviour of an infinite chain of identical qubits
inside a resonator, from the point of view of THz or sub-THz
electromagnetic wave propagation in such a quantum medium. We show
that by placing the qubits in a quantum superposition state, some
interesting possibilities can be realized, including ``breathing''
photonic crystals and an ``Archimedean screw'' transport of 
classical electromagnetic modes. 

For the lack of a better term, we call such qubit structures, considered from the point of view of macroscopic propagation of electromagnetic field, {\em quantum metamaterials}. This because (classical) metamaterials allow additional ways to control the propagation of electromagnetic fields, not available to standard materials. (Alternative approaches to superconducting metamaterials  were investigated in Refs.[\onlinecite{Salehi2007,Salehi2005,Ricci,Lazarides,Wang,Du,Ricci-b}].) Similarly, our proposed quantum metamaterials allow additional ways of controlling the propagation of electromagnetic waves, not possible with normal classical structures. Indeed, the {\it coherent quantum dynamics} of qubits determines the THz ``optical'' properties in the system.

\section{Model} 
\begin{figure}[btp]
\begin{center}
\includegraphics[width=14.0cm]{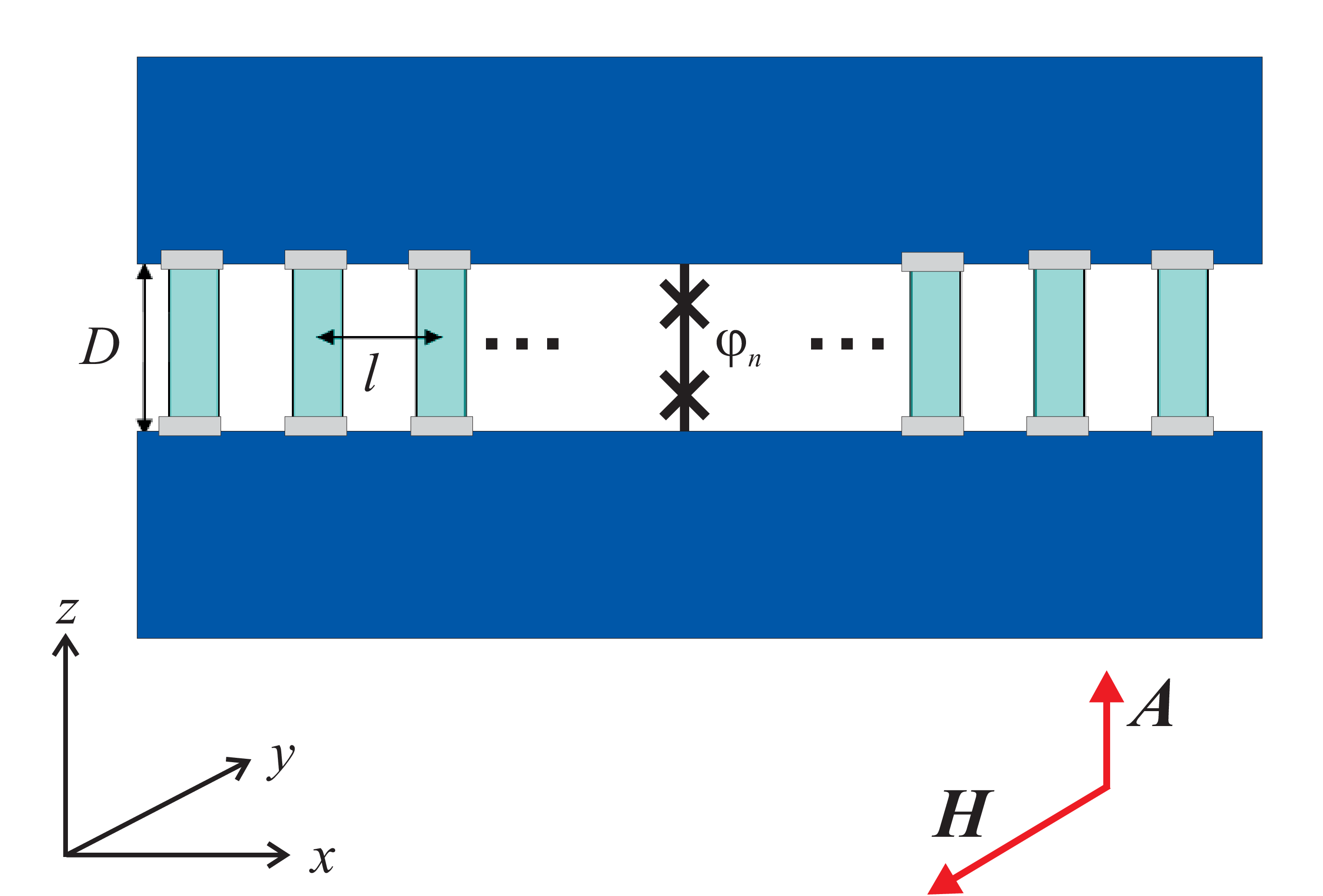}
\end{center}
\caption{(Color online.) Geometry of the system. Identical charge qubits are placed at equal intervals $l$ between bulk superconductors separated by a distance $D$.} \label{fig1}
\end{figure}

As a model, we choose a set of identical charge qubits placed at equal intervals, $l$, between two bulk superconductors separated by a distance $D$ (Fig.~\ref{fig1}). Each qubit is a small superconducting island connected to each superconducting bank by a Josephson junction. The superconducting phase on the $n$th island is $\varphi_n$. When treated  quantum mechanically, such an island indeed consitutes a qubit, if its total capacitance is small enough~\cite{You2005a,Wendin2006}.  The
magnetic field $\textbf{H}$ is applied normal to the structure (in the
$y$ direction) and the vector potential $\textbf{A}$ has  only a $z$-component. We denote by $A_{zn}$ the vector-potential between the $n$-th and
$(n+1)$-th qubits.

The structure of Fig.~\ref{fig1} is a 1D waveguide with
 the energy per unit length
\begin{eqnarray}\label{En}
\nonumber {\cal E}&=&\frac{E_J}{2\omega_J^2}\left[\left(\frac{2\pi
D\dot{A}_{zn}}{\Phi_0}+\dot{\varphi_n}\right)^2+\left(\frac{2\pi
D\dot{A}_{zn}}{\Phi_0}-\dot{\varphi_n}\right)^2\right]\\
&-&E_J\Biggl\{\cos\left[\varphi_n\:+\:\frac{2\pi
DA_{zn}}{\Phi_0}\right]\\
\nonumber\!\!&+&\!\!\!\cos\Biggl[\varphi_n\:-\:\frac{2\pi
DA_{zn}}{\Phi_0}\Biggr]\Biggr\}
+\frac{Dl}{8\pi}\left(\frac{A_{zn+1}\!-\!A_{zn}}{l}\right)^2.
\end{eqnarray}
Here the dot denotes $\partial/\partial t$, the Josephson energy, Josephson frequency, critical current and junction capacity are respectively $E_J=\Phi_0I_c/2\pi c$, $\omega_J=eI_c/\hbar C$, $I_c$ and $C$; $\Phi_0 = h/2e$ is the flux quantum. We took into account that, in the presence of the vector potential, the superconducting phase differences across the junctions of the $n$th qubit, $\pm \varphi_n$, acquire a gauge term,
 $ \alpha_n=2\pi DA_{zn}/\Phi_0.$ Introducing the dimensionless units
$E = {\cal E}/E_J,$ and $\:t\rightarrow\omega_0t,$
we rewrite Eq.~\eqref{En} as
\begin{equation}\label{en}
E=\dot{\varphi}_n^2+\dot{\alpha}_n^2-2\cos\alpha_n\cos\varphi_n+\beta^2(\alpha_{n+1}-\alpha_n)^2,
\end{equation}
where
\begin{equation}\label{beta}
\beta^2=\frac{1}{8\pi Dl E_J}\left(\frac{\Phi_0}{2\pi}\right)^2 \equiv \frac{E_{\rm EM}}{E_J}
\end{equation}
characterizes the ratio $(E_{\rm EM}/E_J)$ of electromagnetic and Josephson energies.

In this paper  the electromagnetic (EM) field is treated as a {\em classical} wave.
We also assume that  its amplitude is small, $\alpha_n\ll 1$. This means that
the magnetic flux per unit cell area $H_yD\times l$ is much
smaller than  $\Phi_0$. Under such assumptions, the
Hamiltonian for a single qubit is
\begin{equation}\label{Ham_JJ}
{\cal H}=-\left(\frac{\partial}{\partial
\varphi_n}\right)^2 - \alpha_n^2\;\cos\varphi_n.
\end{equation}
We restrict the states of each qubit to either its ground state $|0\rangle$, with
energy $E_0$, or  excited state $|1\rangle$ with energy
$E_1$.  This is justified due to the nonlinearity of the Josephson potential.
We are  not  concerned here with  decoherence effects in the qubits, concentrating exclusively on their interaction with the electromagnetic wave in the system. This idealization is justified as long as the decoherence time exceeds the wave propagation time across a significant number of  unit cell periods. This is not unrealistic, given the high quality of superconducting resonator-qubit devices already achieved by recent experiment (e.g., in Ref.~\onlinecite{Wallraff2004} the quality factor $Q$ exceeded $Q = 10^4$ at 6 GHz). 

In the absence of an EM field, the wave function of
the system, $\Psi_n$, is a sum
\begin{equation}\label{psi}
\Psi_n = C_0^{n0} \;|0\rangle\; e^{i\varepsilon t/2}\; + \;C_{1}^{n0}\;
|1\rangle \;e^{-i\varepsilon t/2}\,,
\end{equation}
where $C_k^{n0}$ are constants and $\varepsilon$ is the dimensionless
excitation energy,
$$
\varepsilon=\frac{E_1-E_0}{\hbar\omega_0}.
$$
In the presence of an EM field, the coefficients $C_k^n$ become
time-dependent, and, as it follows from Eq.~\eqref{Ham_JJ}, these obey the
relations~\cite{Landau1995}
\begin{equation}\label{pert}
i\frac{dC_k^n}{dt}=\alpha_n^2\sum_{m=1,2}V^n_{km}(t) \; C^n_m(t)
\end{equation}
with the initial conditions $C_k^n(t=0)=C_k^{n0}$. Here
$$V^n_{km}(t)=\langle k|\cos \varphi_n |m\rangle$$ are matrix
elements of the interacting
field-qubit interaction from \eqref{Ham_JJ}, calculated in the Heisenberg basis $$\left\{\:|0\rangle\exp{(i\varepsilon
t/2)},\:|1\rangle\exp{(-i\varepsilon t/2)} \right\}.$$ Later we will also use
the time-independent matrix elements $V^n_{km}=\langle k|\cos
\varphi_n |m\rangle$ in the basis $\left\{ |0\rangle,\,|1\rangle \right\}$.

Varying the energy \eqref{en}, we obtain the equation for the electromagnetic field in the linear approximation:
\begin{equation}\label{EMF}
\ddot{\alpha}_n-\beta^2\left(\alpha_{n+1}+\alpha_{n-1}-2\alpha_n\right)+\alpha_n\langle
\Psi_n|\cos \varphi_n |\Psi_n\rangle=0.
\end{equation}
The set of Eqs.~\eqref{psi}, \eqref{pert},
and \eqref{EMF} should be supplied by appropriate initial and
boundary conditions.
By controlling the qubits in Eq.~(\ref{pert}), we propose to change the transmission and reflection of EM waves described by Eq.~(\ref{EMF}).

We are interested in the case when the wavelength is large compared to the size of the unit cell. Therefore the
 qubit line can be treated as a
continuous 1D medium with $n\cdot l$ replaced by $x$. The difference equation~\eqref{EMF} for
$\alpha_n(t)$ and $\Psi_n(t)$ is thus replaced by a differential equation
for $\alpha(x,t)$ and $\Psi(x,t)$
\begin{equation}\label{EMF_media}
\ddot{\alpha}-\beta^2\frac{\partial^2\alpha}{\partial
x^2}+V_0\alpha=0,\,\,\, V_0=\langle\Psi(x)|\cos
\varphi(x)|\Psi(x)\rangle
\end{equation}
Within the perturbation theory approach, we present the
electromagnetic wave as a  sum of the larger incident wave,
$\alpha_0$, and a  smaller scattered wave $\alpha_1$. A quantum state
of the system is described by the wave function
\begin{equation}\label{psi_x}
\Psi(x,t)=C_0(x,t)|0\rangle e^{i\varepsilon t/2}+C_1(x,t)|1\rangle
e^{-i\varepsilon t/2}.
\end{equation}
In the unperturbed state the coefficients in this equation are
$C_i=C_i^0(x)$. We present the coefficients $C_i(x,t)$ as a
sum of the unperturbed solution $C_i^0(x)$ and a small perturbation $C_i^1(x,t)$,
$C_i(x,t)=C_i^0(x)+C_i^1(x,t)$, with $|C_i^1|\ll 1$. Using
Eq.~\eqref{pert}, we derive
\begin{eqnarray}\label{c_1}
\nonumber
iC_0^1 &=& \int_0^tdt'\alpha_0^2\left(V_{00}\ C_0^0 + V_{01}\ C_1^0e^{-i\varepsilon t'}\right) \\
iC_1^1 &=& \int_0^tdt'\alpha_0^2\left(V_{11}\ C_1^0+
V_{10}^*\ C_0^0e^{i\varepsilon t'}\right),
\end{eqnarray}
where $V_{ik}=\langle i|\cos\varphi|k\rangle$ are calculated using the unperturbed wave functions,  ~$V^*$ means
complex conjugate of $V$, and $V_{10}^*=V_{01}$.

For the unperturbed EM wave $\alpha_0$, we obtain
from Eq.~\eqref{EMF_media}
\begin{equation}\label{a0}
\ddot{\alpha}_0-\beta^2\frac{\partial^2\alpha_0}{\partial
x^2} + V_0\ \alpha_0=0.
\end{equation}
Here, $(V_0)^{1/2}$ plays the role of the Josephson plasma frequency, which is now controlled by the quantum state and quantum dynamics of the qubits.
For the matrix element $V_0$ we can derive the following expression
\begin{equation}\label{V0}
V_0=|C_0^0|^2\ V_{00}+|C_1^0|^2\ V_{11} + C_0^0C_1^{0*}e^{i\varepsilon
t}\ V_{10} + {\rm h.c.}
\end{equation}
 For simplicity, we assume that
$\alpha_0$ is a standing wave, $\alpha_0=A\cos(\omega
t)\cos[k(\omega) x]$.

\section{Electromagnetic wave propagation through a uniform qubit line}
\subsection{Qubits initially in the ground state $|0\rangle.$}
If {\em all} the qubits are in the ground state $|0\rangle$, then initially
$C_0^0=1$ and $C_1^0=0$. In this case $V_0=V_{00}$ and the wave
vector is
\begin{equation}\label{k(w)}
k(\omega)=\frac{1}{\beta}\sqrt{\omega^2-V_{00}}.
\end{equation}
Thus, the wave can propagate if its frequency  exceeds
$(V_{00})^{1/2}$, which can be interpreted as the ``ground state'' plasma frequency of
the medium. From Eq.~\eqref{c_1} we obtain
\begin{eqnarray}\label{c_io}
\nonumber \frac{C_0^1(x,t)}{V_{00}} &=& -\;\frac{iA^2 \cos^2(kx)}{2}\left\{t+\frac{\sin(2\omega t)}{2\omega}\right\} \\
 \frac{C_1^1(x,t)}{V_{01}} &=&
-\;\frac{A^2 \cos^2(kx)}{2}\Biggl\{\frac{e^{i\varepsilon
t}-1}{\varepsilon} \\ \nonumber
&+&\frac{\varepsilon+e^{i\varepsilon t}\left[2i\omega\sin(2\omega
t)-\varepsilon\cos(2\omega
t)\right]}{4\omega^2-\varepsilon^2}\Biggr\}
\end{eqnarray}
The initial disturbance of the wave function produces a disturbance
$\alpha_1$ in the propagating wave. For this perturbation, using Eq.~\eqref{EMF_media},
we derive
\begin{equation}\label{a1}
\ddot{\alpha}_1-\beta^2\frac{\partial^2\alpha_1}{\partial
x^2}+V_{00}\alpha_1+\Delta V_0\alpha_0=0,
\end{equation}
$\Delta V_0$ being the perturbation of the field-qubit coupling. By
means of Eqs.~\eqref{c_io} we find
\begin{eqnarray}\label{dV}
\nonumber
\Delta V_0(t) &=& -|V_{01}|^2A^2\cos^2(kx) \\
\times\Biggl\{\frac{1}{\varepsilon}&-&\frac{2(2\omega^2-\varepsilon^2)\cos(\varepsilon
t)+\varepsilon^2\cos(2\omega
t)}{\varepsilon\left(4\omega^2-\varepsilon^2\right)}\Biggr\}.
\end{eqnarray}
We see that the electromagnetic wave is in resonance with the qubit line if its
frequency is half the inter-level distance,
$\omega=\varepsilon/2$. This is due to the term proportional to
$\alpha^2$ in the Hamiltonian \eqref{Ham_JJ}. Near the resonance,
the condition $|C_i^1|\ll 1$ is no longer valid and the usual
perturbation approach fails.

\subsection{Qubits initially in the excited state $|1\rangle.$}
If {\it all} qubits are initially in the excited state $|1\rangle$, the solution  is found in  complete analogy to the previous case. As it can
be readily seen, we should only exchange
$0 \leftrightarrow 1$ and $\varepsilon \leftrightarrow
-\varepsilon$ in Eqs.~\eqref{k(w)}-\eqref{dV}
\begin{equation}\label{k(w1)}
k(\omega)=\frac{1}{\beta}\sqrt{\omega^2-V_{11}},
\end{equation}
\begin{eqnarray}\label{c_io1}
\nonumber \frac{C_1^1(x,t)}{V_{11}} &=& -\frac{iA^2 \cos^2(kx)}{2}\left\{t+\frac{\sin(2\omega t)}{2\omega}\right\} \\
 \frac{C_0^1(x,t)}{V_{10}} &=&
-\frac{A^2 \cos^2(kx)}{2}\Biggl\{\frac{1-e^{-i\varepsilon
t}}{\varepsilon} \\ \nonumber
&+&\frac{-\varepsilon+e^{-i\varepsilon
t}\left[2i\omega\sin(2\omega t)+\varepsilon\cos(2\omega
t)\right]}{4\omega^2-\varepsilon^2}\Biggr\},
\end{eqnarray}
and for the electromagnetic wave we obtain
\begin{equation}\label{a11}
\ddot{\alpha}_1-\beta^2\frac{\partial^2\alpha_1}{\partial
x^2}+V_{11}\alpha_1+\Delta V_1\alpha_0=0.
\end{equation}
In Eq.~(\ref{a11})
\begin{equation}\label{dV1}
\nonumber
\Delta V_1(t) = -  \Delta V_0(t),
\end{equation}
where $\Delta V_0(t)$ is given by Eq.~(\ref{dV}).

The excited qubit line is an {\em active medium}, and one should expect a resonance pumping of the electromagnetic wave as it propagates along. However, $\Delta V_1\rightarrow 0$ at
$2\omega\rightarrow\varepsilon$ (and $\Delta V_0\rightarrow 0$ at
$2\omega\rightarrow\varepsilon$ as well). This ``paradox'' only reflects the limitations of  the first order perturbation approximation, where $|C_i^0+C_i^1|^2=|C_i^0|^2$. In other words, to first order, the qubit
energy does not change. To describe
the pumping effect, we must take into account the  higher order terms, which is beyond the scope of our current investigation.

Finally, let $C_0=C_1=1/2$. In this case all qubits {\it `rotate'} between the ground
and excited state. The matrix element in
Eq.~\eqref{EMF_media} is now
\begin{equation}\label{rotaV}
V_0(t) = \frac{1}{4}\left[V_{00}+V_{11}+2V_{01}\cos(\varepsilon
t)\right].
\end{equation}
Let us now assume, for simplicity, that the frequency of the electromagnetic
wave is high, $\omega\gg \varepsilon$. Then its wave vector
is a slowly oscillating function
\begin{equation}\label{k(t)}
k(\omega,t)\approx\sqrt{\omega^2-\frac{V_{00}+V_{11}+2|V_{01}|\cos(\varepsilon
t)}{4\beta^2}}
\end{equation}
If the wave frequency $\omega$ is close to the threshold value,
$$\omega_c=\sqrt{V_{00}+V_{11}}/2\beta,$$ then the qubit line will
alternate between transparent and reflecting state with a frequency
$\varepsilon$, as the wave vector $k(t)$ switches between real and
imaginary values. In addition, the qubit line produces electromagnetic waves with
frequencies $\varepsilon$ and $\omega\pm\varepsilon$.

\section{Electromagnetic wave at resonance}
Near the
resonance, $2\omega -\varepsilon \equiv \Delta\omega\ll\omega$, we can use a resonant perturbation approach~\cite{Landau1995}, instead of a first-order perturbation approximation. In doing
so, we drop all the terms in Eqs.~\eqref{pert} except the resonant
ones. As a result, we have
\begin{eqnarray}\label{reso}
\nonumber
i\dot{C}_0 &=& -\Omega e^{i\Delta\omega t}\ C_1,\\
i\dot{C}_1 &=& -\Omega^* e^{-i\Delta\omega t}\ C_0,
\end{eqnarray}
where $$\Omega(x)=A^2\cos^2[k(\omega)x]V_{01}/4.$$ The solution of
this system is
\begin{eqnarray}\label{CC}
\nonumber C_0(t)\!\! &=&\!\! \frac{e^{i\Delta\omega
t/2}}{\Omega^*}\left[g_1\left(\Lambda-\frac{\Delta\omega}{2}\right)e^{i\Lambda
t}
-g_2\left(\Lambda+\frac{\Delta\omega}{2}\right)e^{-i\Lambda t}\right]\\
C_1(t)\!\! &=&\!\! e^{-i\Delta\omega t/2}\left(g_1e^{i\Lambda
t}+g_2e^{-i\Lambda t}\right),
\end{eqnarray}
where $$\Lambda=\sqrt{\Omega^2+\left(\Delta\omega\right)^2/4}\ ,$$
and $g_1$, $g_2$ are constants. The coefficients
$C_i$ satisfy the normalization condition $|C_0|^2+|C_1|^2=1$. If
at $t=0$ the system was in the ground state, we
obtain
\begin{eqnarray}\label{CGr} \nonumber
C_0(t) &=& e^{i\Delta\omega t/2}\left[\cos(\Lambda t) -\frac{i\Delta\omega}{2\Lambda}\sin(\Lambda t)\right], \\
C_1(t) &=&-\frac{i|\Omega|}{\Lambda}e^{-i\Delta\omega
t/2}\sin{(\Lambda t)}.
\end{eqnarray}
At the resonance $\Delta\omega\!\!=\!\!0$,
$$\Psi\!\!=\!\!\exp{(i\varepsilon t/2)}\cos(|\Omega|
t)|0\rangle\!\!-\!\!i\exp{(-i\varepsilon t/2)}\sin(|\Omega|
t)|1\rangle,$$ and each qubit periodically oscillates between its ground
and excited states. The frequency of these transitions varies with
the qubit position, since $\Omega=\Omega(x)$. It can be considered as a spatially-dependent quantum beat frequency. If the system was
initially in its excited state, then, we have
\begin{eqnarray}\label{Cexc} \nonumber
C_0(t) &=&-\frac{i|\Omega|}{\Lambda}e^{i\Delta\omega
t/2}\sin{(\Lambda t)}, \\
C_1(t) &=& e^{-i\Delta\omega t/2}\left[\cos(\Lambda t)
-\frac{i\Delta\omega}{2\Lambda}\sin(\Lambda t)\right]
\end{eqnarray}
and at $\Delta\omega\!\!=\!\!0$,
$$\Psi\!\!=\!\!-\!i\exp{(i\varepsilon t/2)}\sin(|\Omega|
t)|0\rangle\!\!+\!\!\exp{(-i\varepsilon t/2)}\cos(|\Omega|
t)|1\rangle.$$

Using these expressions for the wave functions, we
find the matrix element $V_0$ in Eq.~\eqref{V0}. Assuming for
simplicity that  $V_{ik}$ are real and
$\Delta\omega^2\ll |\Omega|^2$, we obtain
\begin{eqnarray}\label{Vreso}
\nonumber
  V_0(t) &=& \frac{V_{00}+V_{11}}{2}\pm \frac{V_{00}-V_{11}}{2}\cos{\left(2|\Omega|t\right)} \\
 &-&V_{01}\sin{\left(2|\Omega|t\right)}\sin{\left(\varepsilon t+\Delta\omega\right)},
\end{eqnarray}
where the signs `$+$' and `$-$' correspond to the initial ground and excited
state for qubits, respectively. If the transitions between
the ground and  excited states are suppressed, $|V_{01}|\ll
|V_{00}-V_{11}|$, then, in resonance,  the electromagnetic wave
$\alpha_0$ has a time-dependent wave vector,
$$k(\omega,t)=\sqrt{\omega^2-V_0(t)},$$ where $V_0(t)$ varies
between $V_{00}$ and $V_{11}$. These  results are valid if
the wave-qubit interaction  does not distort it too much:  $|\alpha_1^2|\ll |\alpha_0^2|$. In any case
the condition
\begin{equation}\label{valid}
|\Omega|\ \sim \ A^2|V_{01}|\ \ll \ \omega\ \sim\ \varepsilon
\end{equation}
must be fulfilled.

\section{Quantum metamaterial photonic crystal}

In analogy with  photonic crystals~\cite{PHOTONIC}, the interaction of the electromagnetic wave with qubits can produce a frequency gap   in the
spectrum of the propagating wave, if the qubit states are
periodically modulated in space. For example, suppose the
qubits are in the $|\gamma\rangle$ or $|\delta\rangle$ state with a
spatial period $2L$. The wave obeys the equation
\begin{equation}\label{Photo_a1}
\ddot{\alpha}-\beta^2\alpha_{xx} +
  V_{\gamma\gamma} \ \alpha=0
\end{equation}
or
\begin{equation}\label{Photo_a2}
\ddot{\alpha}-\beta^2\alpha_{xx}+\
  V_{\delta\delta} \ \alpha=0.
\end{equation}
The states $|\gamma\rangle$ or $|\delta\rangle$ can be either
stationary (eigenstates of the qubit Hamiltonian), or their superpositions. In the latter case, the
photonic crystal discussion makes sense only if the quantum beat
frequency is small compared to the frequency of the propagating
wave, that is, $\varepsilon^2\ll |V_{00}|$ and $|V_{11}|$.

\begin{figure}[btp]
\begin{center}
\includegraphics[width=8.0cm]{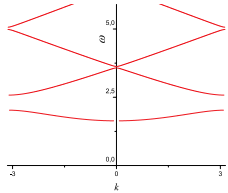}
\end{center}
\caption{ (Color online.) Photonic crystal spectrum obtained from a qubit transmission line. The  $\omega(k)$ (red) shown was calculated 
for a  periodic array of qubit states (ground and excited). The ratio  $V_{00}/V_{11}=5$.}\label{fig2}
\end{figure}

\begin{figure}[btp]
\begin{center}
\includegraphics[width=8.0cm]{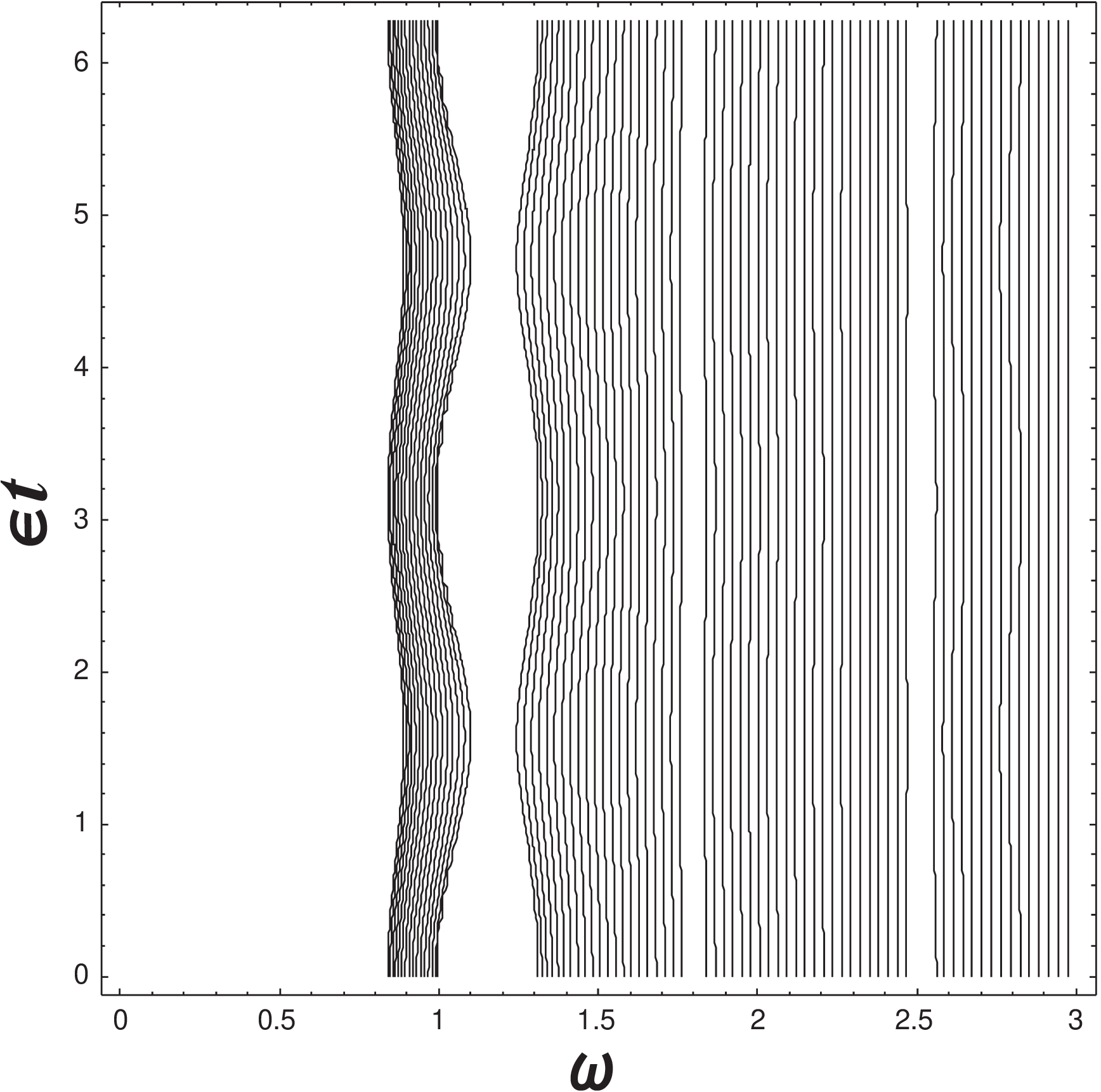} 
\end{center}
\caption{ Breathing photonic crystal: contour curves of the wave vector $k$ as a function of $\omega$ and $\varepsilon t$ in the situation described by Eq.~(\ref{eq_breathing}). 
The parameters used here are $V_{00} = V_{01}=1,$ $V_{11}=2$, $\beta = 0.5$, $L=2$. The time-dependent gaps in the spectrum are clearly seen.}\label{fig2A}
\end{figure}

Following the usual band-theory approach for electrons in a crystal
lattice, we seek the solution of Eq.~(\ref{Photo_a1},\ref{Photo_a2}) in the form
of a Bloch wave $\alpha(t,x)=u(x,k)\exp(ikx-i\omega t)$, where
$u(x,k)$ is a periodic function of $x$ with the period $2L$, and
the dimensionless wave vector $k$ is in the first Brillouin zone,
$-\pi/L<k<\pi/L$. Consider the $j$th elementary cell of our periodic
structure: for $x_j<x<x_j+L$ all the qubits are in
state $|\gamma\rangle$, and for $x_j+L<x<x_j+2L$ in
the state $|\delta\rangle$. In both regions, the solution
$\alpha(t,x)$ of  Eq.~(\ref{Photo_a1},\ref{Photo_a2}) is a sum of
exponential terms multiplied by constants $C_j$. Using the
continuity of $\alpha$ and $\partial\alpha/\partial x$ at the
boundaries of different regions and the periodicity of the Bloch
functions $u(x,k)$, we obtain a set of homogeneous linear
equations for $C_j$. The nontrivial solution of these equations
exists only if the determinant of the set of equations is zero.
Then, after straightforward algebra, we obtain the dispersion
equation for the frequency $\omega(k)$ in the form
\begin{eqnarray}
\nonumber \cos(\kappa_\gamma L)\cos(\kappa_\delta L) &-&
\frac{\kappa^2_\gamma+\kappa^2_\delta}{2\kappa_\gamma\kappa_\delta}\sin(\kappa_\gamma L)\sin(\kappa_\delta L) \\
&=& \cos(2kL),
\end{eqnarray}
where
\begin{equation}\label{kappa}
\kappa^2_\gamma=\frac{\omega^2-V_{\gamma\gamma}}{\beta^2},\qquad
\kappa^2_\delta=\frac{\omega^2-V_{\delta\delta}}{\beta^2}.
\end{equation}
This equation predicts the spectrum $\omega(k)$ with   gaps if
the difference between $\kappa_\gamma$ and $\kappa_\delta$ is large
enough; that is, $|\kappa_\gamma^2-\kappa_\delta^2|\gtrsim 1$, or
\begin{equation}\label{gap}
|V_{\gamma\gamma}-V_{\delta\delta}|\gtrsim\beta^2\,.
\end{equation}
Thus, in order to form a photonic crystal in the qubit line,
the Josephson energy $E_J$ must be large compared to the magnetic energy
or, according to Eq.~\eqref{beta},
\begin{equation}\label{condition}
E_J \gg \frac{1}{8\pi }\left(\frac{\Phi_0}{\pi D\;l}\right)^2.
\end{equation}
The characteristic dependence of $\omega(k)$ is shown in
Fig.~\ref{fig2}. Note that the gap value for the first zone is of the order
of unity if the condition Eq.~\eqref{gap} is valid.

The {\it gap depends on the quantum state of the qubits}, making this a {\it quantum photonic crystal}. Changing the microscopic quantum state of the qubits changes the macroscopic electromagnetic response of the system.

A more interesting situation arises if one or both of the qubit
states are not the eigenstates $|0\rangle$ or $|1\rangle$, and therefore this system would exhibit  {\em quantum beats} between the two.  Let, e.g.,
$|\gamma \rangle = |0\rangle$ and $$|\delta \rangle = \left\{
|0\rangle e^{i\varepsilon t} + |1\rangle e^{-i\varepsilon t}\right\}/2.$$
Then
$V_{\gamma\gamma} = V_{00}$ and
$$V_{\delta\delta}(t)  = [V_{00}+V_{11}+2V_{01}\cos(\varepsilon t)]/4.$$
In this case, the photonic crystal arises if any of the matrix
elements is of the order of unity. The  frequency gap is
modulated by the value of $V_{01}/2$ with the period
$\Delta t=2\pi/\varepsilon$. If $V_{01}\sim 1$, then the
modulation is significant. If 
$$|\gamma \rangle = \left\{|0\rangle
e^{i\varepsilon t} - |1\rangle e^{-i\varepsilon t}\right\}/2$$ and
$$|\delta \rangle = \left\{ |0\rangle e^{i\varepsilon t} + |1\rangle
e^{-i\varepsilon t}\right\}/2,$$ then
\begin{eqnarray} V_{\gamma\gamma}(t) = [V_{00}+V_{11}-2V_{01}\cos(\varepsilon t)]/4, \nonumber\\
\label{eq_breathing}\\
V_{\delta\delta}(t) = [V_{00}+V_{11}+2V_{01}\cos(\varepsilon t)]/4. \nonumber
\end{eqnarray}
In this case the photonic crystal appears if $V_{01}\sim 1$. The
gap is strongly modulated in time from zero, at
$t=(2n+1)\pi/2\varepsilon$, to its maximum value, when
$t=n\pi/\varepsilon$ (here $n$ is an integer). Thus we obtain an interesting uniformly {\it ``breathing''} photonic crystal, shown in Fig.\ref{fig2A}.

\section{Quantum `Archimedean screw'}
Let an external source produce a slow ``control wave'' propagating along
the qubit line. That is, let the coefficients $C_i$ in the wave function
be
\begin{equation}\label{C-W}
C_0(x,t) = \sin{\left(\omega_0t-k_0x\right)},\quad
C_1(x,t) = \cos{\left(\omega_0t-k_0x\right)},
\end{equation}
where $\omega_0$ and $k_0$ are the frequency and the wave vector
of the control wave, respectively, and both these values are small
compared with the electromagnetic wave frequency and wave vector.
Such a wave can be produced, e.g., by applying to the qubits a RF signal with a position-dependent phase, and inducing Rabi oscillations between their ground and excited states.

The matrix element $V_0$ in the wave equations then takes
the form
\begin{eqnarray}\label{V-R}
\nonumber
V_0(x,t) &=& \frac{V_{00}+V_{11}}{2}-\frac{V_{00}-V_{11}}{2}\cos{\left[2\left(\omega_0t-k_0x\right)\right]} \\
&+&V_{01}\sin{\left[2\left(\omega_0t-k_0x\right)\right]}\cos{\varepsilon
t},
\end{eqnarray}
where we now assume that $V_{01}$ is real. To simplify the
problem, we now assume that $|V_{01}|\ll |V_{00}\pm V_{11}|$ and the
last term, which describes the qubit relaxation, can be neglected.
Thus, $V_0=V_0(\xi)$, where $\xi=\omega_0t-k_0x$.

\begin{figure}[btp]
\begin{center}
\includegraphics[width=8.0cm]{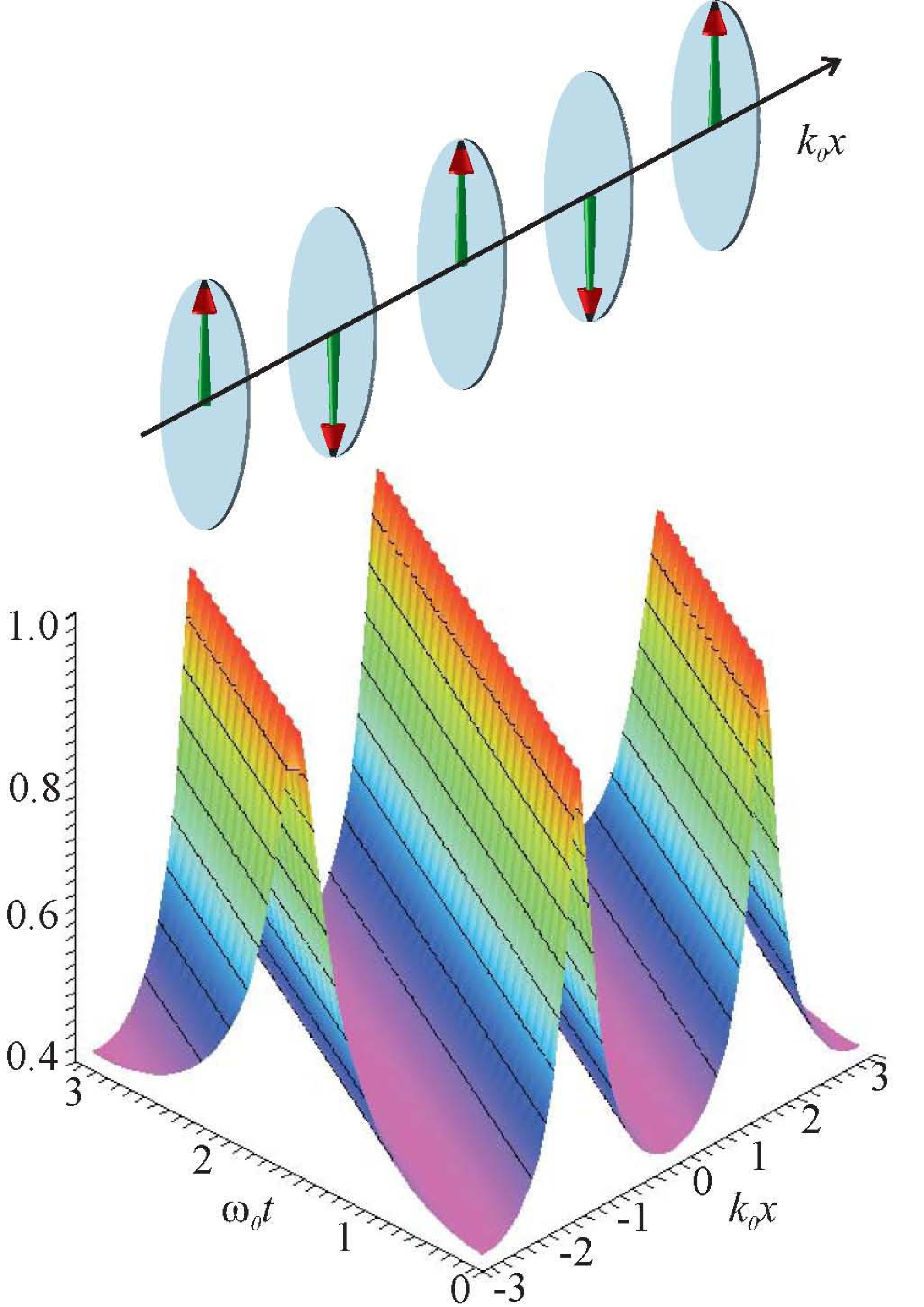}
\end{center}
\caption{ (Color online.) Quantum `Archimedean screw': (top) schematic diagram of how the wave amplitude $A(x,t)/A_0$ varies with position at a given moment in time; (bottom) the dependence of the wave amplitude on time and position;
$V_{00}+V_{11}=2$, $V_{00}-V_{11}=1$, $\beta=0.15$, $k_0=0.1$,
$\omega_0=10^{-3}$, $\omega=1.23$.
 }\label{fig3}
\end{figure}

The function $V_0(\xi)$ varies slowly and we can seek the solution
of Eq.~\eqref{EMF_media} in the form
\begin{equation}\label{a-EMR}
\alpha(t,\xi)=A(\xi)\;\sin{\!\left[\omega_0t-\eta(\xi)\right]},
\end{equation}
where the wave amplitude $A(\xi)$ varies slowly and we can neglect
terms with $A''(\xi)$ (here $f'\; \equiv \;df/d\xi$). Substituting Eq.~\eqref{a-EMR} into Eq.~\eqref{EMF_media}
and separating terms with
$\sin{\!\left[\omega_0t-\eta(\xi)\right]}$ and
$\cos{\!\left[\omega_0t-\eta(\xi)\right]}$, we derive two
equations for the wave amplitude $A$ and the phase $\eta$ in the
form
\begin{eqnarray}\label{A-xi}
\nonumber
  \frac{2A'}{A} &=&-\frac{\left(\beta^2k_0^2-\omega^2_0\right)\eta''}{\left(\beta^2k_0^2-\omega^2_0\right)\eta'+\omega\omega_0}, \\
  \left(\eta'\right)^2&+&2\eta'\frac{\omega\omega_0}{\beta^2k_0^2-\omega^2_0}+\frac{V_0-\omega^2}{\beta^2k_0^2-\omega^2_0} =
  0.
\end{eqnarray}
These equations are valid if
\begin{eqnarray}\label{valid2}
\left|\frac{A''}{A}\right|\ll
\left(\eta'\right)^2; \\ \left|\frac{A''}{A}\right|\ll \frac{\left(\omega-\omega_0\eta'\right)^2}{\left(\omega_0\eta'\right)^2}. \nonumber
\end{eqnarray}
Integrating the first of these equations we get
\begin{equation}\label{AR}
2\ln{A}=-\ln{\left(\eta'+\frac{\omega\omega_0}{\beta^2k_0^2-\omega^2_0}\right)}+{\rm const.}
\end{equation}
The second equation is a quadratic equation for $\eta'$. Choosing
its positive root and substituting it into Eq.~\eqref{AR} we get
\begin{equation}\label{eta'}
\frac{A}{A_0}=\left[\frac{\left(\beta^2k_0^2-\omega^2_0\right)^2}{\beta^2k_0^2\omega^2-\left(\beta^2k_0^2-\omega_0^2\right)V_0}\right]^{1/4},
\end{equation}
where $A_0$ is a constant defined by the boundary conditions. The
dependence $A(x,t)/A_0$ is shown in Fig.~\ref{fig3} for
different values of parameters chosen within the validity range,
Eq.~\eqref{valid2}. It is easy to see that  the
wave amplitude $A(t)$ achieves its maximum if the denominator in
Eq.~\eqref{eta'} becomes small at some moments in time. The value
$\beta^2k_0^2\omega^2-\left(\beta^2k_0^2-\omega_0^2\right)V_0$
should be positive. Assuming that $V_{00}>V_{11}$, we can write the
last condition in the form
\begin{equation}\label{cond}
\omega^2>\omega_c^2=V_{11}\left(1-\frac{\omega_0^2}{\beta^2k_0^2}\right).
\end{equation}
The variation of $A(x,t)/A_0$ is maximum if $\omega$ is close to
$\omega_c$; however, the conditions in Eq.~\eqref{valid2} should be
fulfilled.

Returning to Fig.~\ref{fig3}, we see that the maxima of fast-oscillating electromagnetic field are transferred through the system at a pace, set by the much slower control frequency, $\omega_0$, reminding of the Archimedean screw, or a meat grinder. 

\section{Conclusions} Here we show that, for a classical electromagnetic wave, a line of qubits inside a superconducting cavity plays the role of a 1D transmission line with interesting characteristics and rich new physics. In particular, the quantum superposision of qubit states produces a ``breathing'' state with   transparency changing with the quantum beat frequency of a single qubit. More interestingly, a periodic arrangement of qubit states yields a quantum
photonic crystal, which can also be put into a breathing mode. A time-domain control of the qubits allows to realize an ``Archimedean screw'' state, where the incident electromagnetic wave is periodically modulated, and the regions of its maximum amplitude are carried along the qubit line with a desired speed. The investigation of the action of this system as an active medium requires a special treatment beyond the lowest-order perturbation theory, and will be the subject of future research.

\acknowledgements

This work was supported in part by the National Security Agency (NSA), Laboratory Physical Science (LPS), Army Research Office (ARO), and National Science Foundation (NSF) Grant No. EIA-0130380, JSPS-RFBR 06-02-91200.
A.Z. is grateful to J. Young for fruitful discussions and was partially supported by the Natural Sciences and Engineering Research Council of Canada (NSERC) Discovery Grants Program. S.S. acknowledes support from the Ministry of Science, Culture and Sport of Japan via the Grant-in Aid for Young Scientist No. 1874002, the EPSRC via ARF. No. EP/D072581/1, and ESF network-programme ``Arrays of Quantum Dots and Josephson Junctions.''

\end{document}